\documentclass[aps,prx,reprint,superscriptaddress,longbibliography]{revtex4-2}
\usepackage{graphicx}
\graphicspath{ {./images_entropy/} }

\usepackage[normalem]{ulem}
\usepackage{bm} 
\usepackage[colorlinks=true, citecolor=blue, linkcolor=blue, urlcolor=blue]{hyperref}
\usepackage{pinlabel} 

\usepackage{amsmath}
\usepackage{algorithm}
\usepackage{algpseudocode}
\usepackage{float} 
\usepackage{natbib}

\usepackage{lipsum} 
\usepackage{physics}
\usepackage{amssymb}
\usepackage{mathrsfs}

\usepackage{xcolor} 
\usepackage[T1]{fontenc}

\newcommand{\blu}{\textcolor{blue}}

\newcommand{\wdr}{\omega_{\rm dr}}
\newcommand{\Icoh}{\mathcal{I}_{\rm coh}}
\newcommand{\Iincoh}{\mathcal{I}_{\rm inc}}
\definecolor{mypink}{RGB}{255,0,255}

\usepackage{color}

\begin{document}

	\title{Entropy Transport in Programmable Quantum Junctions}

	\author{Radhika Joshi}
	\affiliation{Peter Gr\"unberg Institute for  Quantum Computing Analytics (PGI-12), Forschungszentrum J\"ulich, J\"ulich 52425, Germany}
	\affiliation{Institute for Quantum Information, RWTH Aachen University, D-52056 Aachen, Germany}

     \author{Yuli V. Nazarov}
	 \affiliation{Kavli Institute of Nanoscience, Delft University of Technology, P.O. Box 5046, 2600GA Delft, The Netherlands}
 
	 \author{Mohammad H. Ansari}
	\affiliation{Peter Gr\"unberg Institute for  Quantum Computing Analytics (PGI-12), Forschungszentrum J\"ulich, J\"ulich 52425, Germany}
	\affiliation{Institute for Quantum Information, RWTH Aachen University, D-52056 Aachen, Germany}
	
\begin{abstract}
\noindent
We show that driven qubit junctions enable programmable control of physical entropy transport, with entropy conductance governed by quantum dynamics rather than by reservoir parameters alone. By comparing two simple quantum architectures---a driven single-qubit junction and a driven two-qubit junction---we find that the two-qubit junction enhances entropy transfer while requiring substantially lower driving power than its single-qubit counterpart. We further reveal two non-intuitive effects in both junctions: a sizable coherent contribution to the entropy current that emerges only under resonant driving, and negative differential entropy conductance, where increasing the thermal bias suppresses entropy flow into the probe reservoir. These results identify quantum logic architectures as programmable devices for entropy transport and suggest routes toward quantum feedback control, reservoir protection and refrigeration in driven quantum circuits.
\end{abstract}

\maketitle

\section{Introduction}

In open quantum systems, a set of quantum degrees of freedom is coupled to external reservoirs, enabling the exchange of energy, information and entropy \cite{cohen,open_Breuer}. Most of the traditional literature has focused on the energy balance of the central quantum system itself \cite{heat_current1,heat_current2,Weimer_2008}. Here, instead, we adopt a transport perspective: the quantum system is treated marely as a junction that mediates the flow of physical quantities between reservoirs. In this setting, reservoir currents of energy, spin or charge are central observables, determined by the microscopic structure and coherent dynamics of the junction \cite{book_transport}.

Entropy occupies a more subtle position. It lies at the interface of quantum thermodynamics, information theory and many-body physics. We use the information-theoretic definition of entropy. The rank-$\alpha$ R\'enyi entropy flow (or entropy conductance) into a reservoir is defined as
\begin{equation}
\mathcal{I}_{\alpha}
\equiv
\frac{dS_{\alpha}}{dt}(\rho)
=
-\frac{d}{dt}
\ln {\rm Tr}\!\left[\rho^\alpha\right],
\label{eq. def Rcond}
\end{equation}
where $\rho$ denotes the reduced density matrix of the reservoir. In contrast to charge, spin, or energy currents, $\mathcal{I}_{\alpha}$ is intrinsically nonlinear in $\rho$ and therefore cannot, in general, be expressed as the expectation value of any conventional quantum-mechanical operator.
 This fundamental nonlinearity places entropy transport outside the scope of standard transport theory and renders its direct evaluation in macroscopic, infinite-dimensional reservoirs a highly nontrivial problem. 
 Nevertheless, the entropy current is a fundamental measure of irreversibility, information loss, correlation flow and the redistribution of entanglement \cite{nielsen02quantum,Crooks,time_entropy,Jaynes_entropy}. 

 Only recently has a mathematically consistent framework been developed to address this challenge by extending nonequilibrium quantum thermodynamics to nonlinear information measures, which enables a perturbative treatment of entropy transport \cite{renyi_Yuli,KAN_method}. The physically relevant von Neumann entropy flow follows through the analytic continuation  $\lim_{\alpha \to 1} d \mathcal{I}_\alpha/d\alpha$, \cite{renyi_Yuli}. 
 Previous studies have primarily examined entropy flow in driven single qubit, harmonic oscillators and a three-level photovoltaic cell \cite{renyi_heatEngine, Ansari_2017e, entropy}.
 Remarkably, applying this formalism on a driven qubit, serving as a junction of several thermal baths revealed phenomena that have no counterpart in conventional thermodynamics. For example, a pronounced suppression of both R\'enyi and von Neumann entropy currents into a probe reservoir was observed,  demonstrating that quantum coherence can fundamentally reshape the transport of information and entropy beyond classical thermodynamic expectations \cite{renyi_heatEngine}. 

 Here we extend the theory from a single two-level qubit junction to a programmable multi-qubit junction operating far from equilibrium. We derive a general expression for the entropy conductance into a weakly coupled probe reservoir attached to $N$-qubit quantum junction with their  dissipative dynamics being governed by $M$ thermal environments with arbitrary coupling strengths. This may be accessed experimentally through the correspondence between entropy flow and the full counting statistics of energy exchange in the reservoir \cite{exact_correspondence}. The resulting entropy conductance can be  programmed by the microscopic junction Hamiltonian, the driving fields and reservoir couplings of the junction.  
 Throughout this work, we use the terms entropy flow, entropy current and entropy conductance interchangeably when referring to the rate at which entropy is transferred into a reservoir.

 As a concrete illustration, we analyse and compare the von Neumann entropy conductance, $\mathcal{I}$, in two paradigmatic settings: a resonantly driven single qubit and a two-qubit junction in which first qubit is driven at the dressed frequency of the second qubit. These setups are analogues of X-gate and Cross-Resonance(CR) gate architectures in quantum-computation \cite{single_qubit_gate,CR_effect,CR_exp,Josephson_fzj,zz_freedom,CR_optimized,ku2020suppression}. For a single qubit, resonant driving generates coherent population inversion, which leaves distinct imprints on the non-equilibrium entropy conductance into the probe reservoir. The two-qubit case is qualitatively different. There, driving one qubit at the dressed frequency of the other induces an effective $ZX$ interaction. This is the the physical mechanism underlying high-fidelity entangling operations in quantum processors \cite{CR_exp}. Through this interaction, the state of the driven qubit governs the excitation dynamics of the second qubit, thereby realizing a directed transfer of quantum information. By embedding these operations between thermal reservoirs, we validate that the entropy conductance acquires characteristic signatures of the conditional information transfer. 

 We demonstrate that a driven two-qubit junction can achieve the same entropy conductance as a single-qubit device at significantly reduced driving power, revealing an intrinsic thermodynamic advantage of multi-qubit architectures. More importantly, entropy transport becomes a programmable phenomenon: coherent driving operations allow the flow of both von Neumann and R\'enyi entropy to be selectively accelerated, suppressed, or redirected through the junction. By tuning the underlying Hamiltonian and dissipation pathways, one can navigate between regimes optimized for low entropy production, efficient information transfer, or enhanced correlation and entanglement generation. This establishes entropy conductance as a controllable design parameter of quantum processors, linking quantum-device engineering directly to thermodynamic performance. Such control is relevant both for minimizing dissipation in quantum thermal machines \cite{masers_heatEngine} and for exploiting entropy generation as a signature of information processing and entanglement formation \cite{nielsen02quantum}.

We also uncover some non-intuitive transport effects that can be useful in developing quantum technologies. In particular, we observe negative differential entropy conductance, the entropy-flow analogue of negative differential thermal conductance, a phenomenon previously studied in engineered non-equilibrium systems and thermal devices \cite{NDTC,NDTC_engineered,NDTC_molecular,diode,transistor}. While many thermal switching mechanisms rely on reversing a temperature gradient, the effect can also take place at fixed reservoir temperatures by controlling the external drive amplitude ($\Omega$) \cite{NDTC_drive}. Since these devices are extensively characterized in superconducting-qubit platforms,\cite{single_qubit_gate,CR_effect,CR_exp,Josephson_fzj,zz_freedom,CR_optimized,ku2020suppression}, the implementations of our theory are attainable.

The article is structured as follows. In Sec.~\ref{sec:general}, we introduce the general multi-qubit model and the corresponding Hamiltonian. In Sec.~\ref{sec:results}, we give general expressions for the Lindblad master equation and the entropy flow into the probe reservoir. We also compare entropy transport in two qubit junction with a single-qubit junction. In Sec.~\ref{sec:discussion}, we discuss the physical effects, including multiple resonances and negative differential entropy conductance. We conclude in Sec.~\ref{sec:conclusion}.

\section{Model}\label{sec:general}

\begin{figure*}[ht]
    \includegraphics[width=0.7\linewidth]{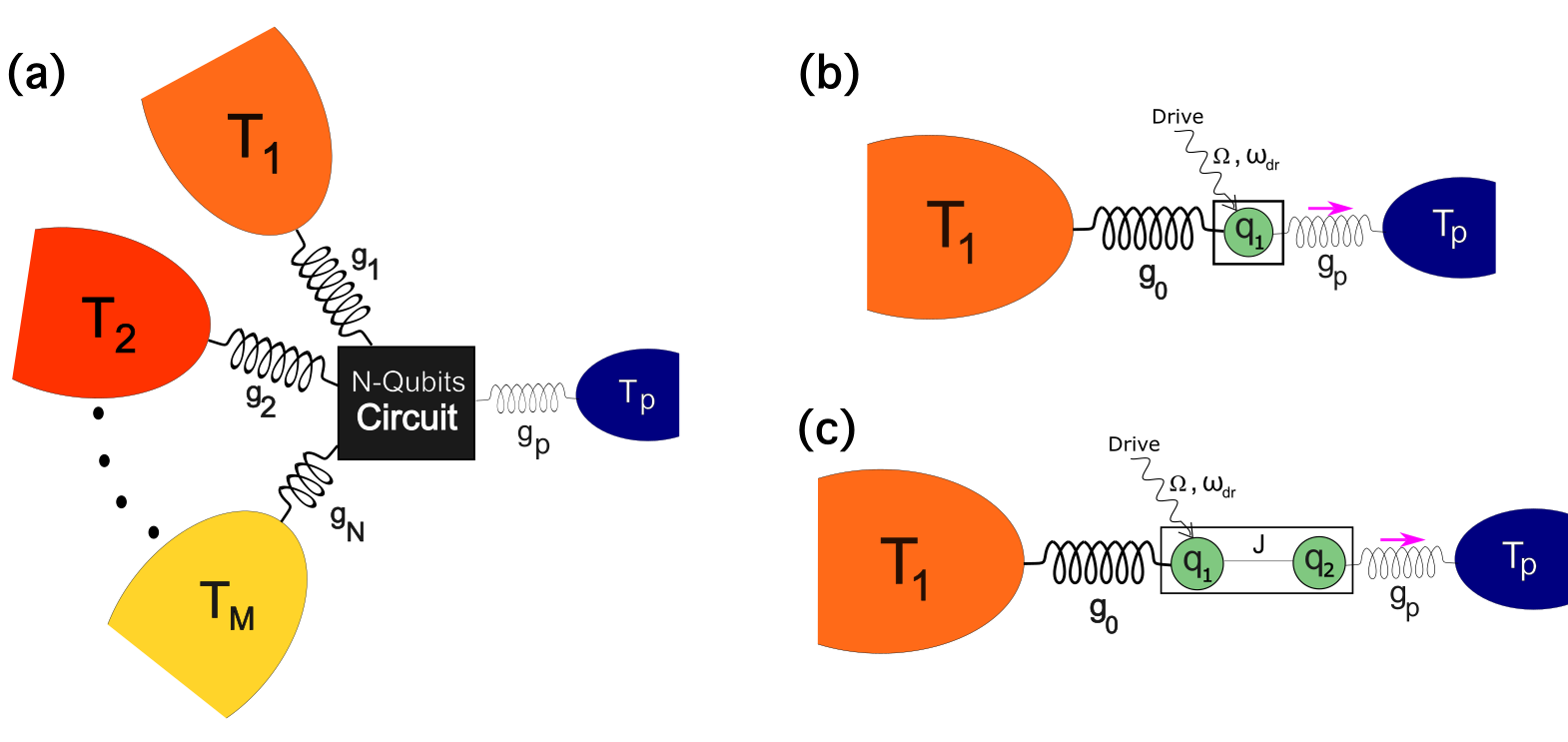}
    \put(-60,125){\color{mypink} {$\mathcal{I}$}}
    \put(-44,50){\color{mypink} {$\mathcal{I}$}}
    \caption{\textbf{(a) General setup}: The quantum junction consists of N-qubits. The qubits can be individually driven and can also interact with each other. The dissipative dynamics of the quantum system are governed by M-thermal reservoirs maintained at temperatures $T_1,\dots ,T_M$. Desired quantity is the entropy flow to the probe reservoir $R_p$, which is kept at temperature $T_p$. \textbf{(b,c) Entropy transport in driven one- and two-qubit junctions}:
    The entropy conductance $\mathcal{I}$ into the probe reservoir
    $R_p$ is evaluated for two representative quantum junctions.
    Reservoirs $R_1$ and $R_p$ are maintained at temperatures $T_1$ and
    $T_p$, respectively.
    \textbf{(b)} A single qubit is coupled to both reservoirs and driven
    resonantly with amplitude $\Omega$.
    \textbf{(c)} A two-qubit junction in which the control qubit $q_1$ is
    driven at the dressed transition frequency $\tilde{\omega}_2$ of the
    target qubit $q_2$. The qubits interact with exchange strength $J$. Qubit
   $q_1$ is coupled to reservoir $R_1$ and qubit $q_2$ is coupled to the probe
    reservoir $R_p$.}
    \label{fig:f_setup}
\end{figure*}
We consider a quantum system composed of $N$ qubits, denoted by $q_i$ with $i=1,2,\cdots,N$, coupled to $M$ thermal reservoirs, as sketched in Fig.~\ref{fig:f_setup}\blu{(a)}. These reservoirs, referred to as the main reservoirs, are each maintained at a fixed temperature and thereby govern the dissipative dynamics of the quantum system. To quantify entropy transport, we additionally introduce a probe reservoir, weakly coupled to the quantum system and kept at nearly zero temperature. This weak-coupling condition ensures that the probe reservoir does not appreciably perturb the intrinsic dynamics of the junction.


The total Hamiltonian is written as
\begin{equation}
H_{\rm tot}=H_S+H_B+H_{SB},
\label{eq:Htot}
\end{equation}
where $H_S$ denotes the Hamiltonian of the quantum system, $H_B$ describes the collection of heat baths (thermal reservoirs), and $H_{SB}$ accounts for the system--bath interaction. 

The quantum system consists of pairwise interacting qubits subject, in general, to external drives. Its unitary dynamics is governed by (from now on, $\hbar = 1$, $k_B=1$)
\begin{equation}
H_S=\sum_i^N H_{q_i}+\sum_{i<j}H_{q_iq_j}+\sum_i^N H_{\mathrm{dr}_i}.
\label{eq:HS}
\end{equation}
The Hamiltonian associated to qubit $q_i$ is
\begin{equation}
H_{q_i}=-\frac{\omega_i}{2}\hat{\sigma}^z_i,
\end{equation}
where $\omega_i$ is the transition frequency of the $i$-th qubit. The pairwise qubit--qubit interaction is taken to be
\begin{equation}
H_{q_iq_j}
=
g_{ij}
\left(\hat{\sigma}^+_i+\hat{\sigma}^-_i\right)
\left(\hat{\sigma}^+_j+\hat{\sigma}^-_j\right),
\end{equation}
where $\hat{\sigma}^{\pm}_i$ are the raising and lowering operators of qubit $q_i$ and $g_{ij}$ denotes the interaction strength between qubits $q_i$ and $q_j$. The external drive applied on qubit $q_i$, is a periodic field of amplitude $\Omega_i$ and frequency $\omega_{\mathrm{dr}_i}$ adds the standard driven-qubit Hamiltonian term \cite{scully1997quantum}
\begin{equation}
H_{\mathrm{dr}_i}
=
\Omega_i\cos(\omega_{\mathrm{dr}_i}t)\,\hat{\sigma}^x_i .
\end{equation}

The environment consists of $M$ heat baths, each of which may exchange energy with one or more qubits. We label the baths by $A=1,2,\cdots,M$. Each bath is modeled as a continuum of harmonic modes,
\begin{equation}
H_B=
\sum_{A=1}^{M}\sum_{ k_A}
\omega_{ k_A}\,
\hat{b}_{ k_A}^{\dagger}\hat{b}_{ k_A},
\end{equation}
where $\hat{b}_{ k_A}$ and $\hat{b}_{ k_A}^{\dagger}$ are the annihilation and creation operators of the bath mode $ k_A$ with frequency $\omega_{ k_A}$. This oscillator-bath representation is the standard microscopic description of a thermal reservoir and applies, for example, to multiple-particle entangled environments \cite{khoshnegar2014toward}, noisy channels \cite{NoiseDWaveQA,pazem2025error}, Ohmic resistors \cite{101038ncomms129642016NatureCommunicationsYanOliverThefluxqubitrevisitedtoenhancecoherenceandreproducibility,ansari2013effect}, nanowire reservoirs \cite{jafari-salim2016stimulated-6f9,jiang2024concurrent}, and superconducting junctions \cite{ansari2011noise-ac0,ansari2013effect}.

The system--bath interaction is assumed to be time independent and bilinear in system and reservoir operators. If bath $A$ couples to qubit $q_i$, the interaction Hamiltonian can be written as
\begin{align}
H_{SB}
=&
\sum_{i=1}^{N}\sum_{A=1}^{M}
\sum_{ k_A}
\left(
g^{-}_{i k_A}\hat{\sigma}^{-}_i\hat{b}_{ k_A}
+
g^{+}_{i k_A}\hat{\sigma}^{+}_i\hat{b}_{ k_A}
\right)
+H.c.
\end{align}
Here $g^{+}_{i k_A}$ and $g^{-}_{i k_A}$ denote the coupling strengths between the raising and lowering operators of qubit $q_i$ and the corresponding annihilation operators of mode $ k_A$ in the bath $A$.

Throughout this work, we focus on cyclic processes governed by dimensionless periodic functions co-rotating with the external drive. This choice enables a transformation to a rotating frame in which the driven Hamiltonian becomes time independent.

\section{Results}\label{sec:results}
We first state the result in its most general form. We consider an $N$-qubit junction mediating transport among $M+1$ macroscopic heat reservoirs held at fixed temperatures. One of these reservoirs, taken to be among the coldest and weakly coupled to the junction, is designated as the probe reservoir. Our central object is the entropy flow into this probe reservoir.

The probe reservoir is not directly coupled to the remaining reservoirs; rather, all inter-reservoir correlations are generated through the quantum junction. Consequently, the non-equilibrium correlations established in the probe are governed by the state of the $N$-qubit system, whose dynamics is obtained using the open quantum system formalism \cite{cohen}. Under the Born-Markov and secular approximations, the dynamics of the system reduces to solving the Lindblad master equation,
\begin{align}\label{eq:L_general}
&\dot{\rho}_{S}(t)
= -i\left[H_{S},{\rho}_{S}(t)\right]\nonumber\\
&+\sum_{i=1}^{N}\sum_{A=1}^{M}\sum_{l=\downarrow,\uparrow}
\Gamma_{iA}^{l}
\left(
L_{iA}^{l}{\rho}_{S}(t)L_{iA}^{l\dagger}
-\frac{1}{2}\left\{L_{iA}^{l\dagger}L_{iA}^{l},{\rho}_{S}(t)\right\}
\right)
\end{align}
where $\rho_S(t)$ is the system density matrix. The dissipative part is expressed in terms of jump operators $L_{iA}^{\downarrow}$ and $L_{iA}^{\uparrow}$, describing relaxation and excitation processes of the $i$-th qubit induced by reservoir $A$. For a two-level qubit, these operators may be taken, for the chosen class of Bath-system interaction $H_{SB}$, as
\begin{align}
    L_{iA}^{\downarrow}=\sigma_i^-,
    \qquad
    L_{iA}^{\uparrow}=\sigma_i^+ .
\end{align}
The corresponding transition rates are
\begin{align}
    \Gamma_{iA}^{\downarrow}
    &= \Gamma_{iA}
    \left[
    n_B\!\left(\frac{\omega_i}{T_A}\right)+1
    \right],
    \\
    \Gamma_{iA}^{\uparrow}
    &= \Gamma_{iA}\,
    n_B\!\left(\frac{\omega_i}{T_A}\right),
\end{align}
where $T_A$ is the temperature of reservoir $A$, $\omega_i$ is the transition frequency of qubit $i$, and
\begin{align}
    \Gamma_{iA}=2\pi |g_{iA}^{+}|^2
\end{align}
sets the reservoir-induced coupling scale, where $g_{iA}=g_{i{k}_A}$ for which $\omega_{k_A}=\omega_i$. Here
\begin{align}
    n_B(\theta)=\frac{1}{e^{\theta}-1}
\end{align}
is the Bose--Einstein occupation factor. These rates determine the evolution of the system density matrix $\rho_S(t)$. The validity of this description requires that the choice of parameters lie in the appropriate parameter regime associated with the Born--Markov, secular, and local-master-equation approximations \cite{local_global,bridging_local_global,Born_Markov}.
\subsection{Entropy conductance}\label{sec:entropyFlow}
The R\'enyi entropy conductance of rank $\alpha$ into the probe reservoir is defined in Eq.~\eqref{eq. def Rcond} in terms of the reduced density matrix of the probe. In the following, we suppress the probe index whenever no ambiguity arises, and all entropy flows are understood to refer to the probe reservoir unless explicitly stated otherwise.

The derivation of the entropy conductance is presented in
Appendix~\ref{app:entropy}. Evaluating the entropy flow with the Keldysh formalism and assuming that the probe is weakly coupled compared to the other reservoirs, $\Gamma_p< \Gamma_A$ for $A\neq p$, we obtain 
\begin{align}
    \mathcal{I}_{\alpha}
    =
    -h(\alpha)
    \left(
    \mathcal{I}_{\rm inc}
    +
    \mathcal{I}_{\rm coh}
    \right),
    \label{eq:renyi}
\end{align}
where the entropy conductance separates into an incoherent contribution
\begin{align}
    \mathcal{I}_{\rm inc}
    =
    \Gamma_{p}^{\uparrow}
    \Tr_S\!\left(
    \sigma_p^{\dagger}\sigma_p \rho_S
    \right)
    -
    \Gamma_{p}^{\downarrow}
    \Tr_S\!\left(
    \sigma_p\sigma_p^{\dagger} \rho_S
    \right),
    \label{eq:same-renyi}
\end{align}
and a coherent contribution
\begin{align}
    \mathcal{I}_{\rm coh}
    =
    \Gamma_p
    \Tr_S\!\left(
    \sigma_p \rho_S
    \right)
    \Tr_S\!\left(
    \sigma_p^{\dagger} \rho_S
    \right).
    \label{eq:cross-renyi}
\end{align}
The universal prefactor is
\begin{align}
    h(\alpha)
    =
    \frac{
    \alpha\, n_B(\alpha\theta_p)
    }{
    n_B(\theta_p)\,
    n_B[(\alpha-1)\theta_p]
    }
    \quad\text{and}\quad
    \theta_p=\frac{\omega_p}{T_p}.
\end{align}

The von Neumann entropy conductance, which will be denoted by $\mathcal{I}$, follows by analytic continuation of the R\'enyi entropy $\lim_{\alpha \to 1} d \mathcal{I}_\alpha/d\alpha$, \cite{renyi_Yuli}, yielding
\begin{align}
    \mathcal{I}
    =
    -\theta_p
    \left(
    \mathcal{I}_{\rm inc}
    +
    \mathcal{I}_{\rm coh}
    \right).
    \label{eq:von}
\end{align}

Within the Keldysh
framework, the total entropy conductance naturally separates into two distinct contributions.
Diagrammatically, as shown in Appendix~\ref{app:entropy}, $\mathcal{I}_{\rm inc}$ originates from single-world
processes, whereas $\mathcal{I}_{\rm coh}$ arises exclusively from
cross-world correlations between replicated Keldysh contours. 
The latter constitutes a genuinely quantum contribution to entropy
transport since it vanishes in the absence of coherent driving within the junction.

The explicit entropy formulae in Eqs.(\ref{eq:same-renyi},\ref{eq:cross-renyi}) reveal that the entropy transport into a reservoir is
not solely determined by the thermodynamic properties of the reservoir
itself. Instead, it depends explicitly on the quantum state of the
intervening circuit through its reduced density matrix $\rho_S$. Entropy
flow therefore becomes a programmable quantity: by manipulating the circuit parameters, one can
actively engineer the transfer of entropy in non-equilibrium quantum
devices.
\subsection{Illustrative examples}
To illustrate the implications of the general formalism developed in the last section, we examine two
paradigmatic quantum-circuit primitives: a driven single-qubit junction and a driven two-qubit junction, as shown in
Fig.\ref{fig:f_setup}\blu{b} and Fig.\ref{fig:f_setup}\blu{c} respectively. These systems represent elementary
building blocks of superconducting quantum processors, namely the X-gate and CR-gate setups and therefore  provide a
minimal platform for exploring the interplay between junction dynamics and
entropy transport.


In general, each qubit may be coupled to multiple thermal reservoirs. The calculations done here extend directly to this situation by
summing the transition rates associated with all reservoirs coupled to a
given qubit. For simplicity and without loss of generality, we restrict the discussion to two reservoirs. The junction dynamics are described by periodically driven Hamiltonian,
\begin{align}
\tilde{H}^{\rm 1q}
&=
-\frac{\omega_0}{2}\hat{\sigma}^z
+
\Omega \cos(\omega_{\rm dr}t)\,\hat{\sigma}^x,
\label{eq:1qH}
\\
\tilde{H}^{\rm 2q}
&=
-\frac{\omega_1}{2}\hat{\sigma}^z_1
-
\frac{\omega_2}{2}\hat{\sigma}^z_2
+
\Omega\cos(\omega_{\rm dr}t)\,\hat{\sigma}^x_1
\nonumber\\
&+J\left(
\sigma_1^\dagger\sigma_2+\mathrm{H.c.}
\right),
\label{eq:2qH}
\end{align}
for the single- and two-qubit junctions, respectively. Here $\Omega$
denotes the drive amplitude and $J=g_{12}$ the inter-qubit exchange coupling. For the rest of this article we fix $\omega_0=\omega_1=5\,\text{GHz}$.
In the two-qubit architecture only the control qubit is driven
externally. Transforming into the frame rotating with the drive frequency
and applying the rotating-wave approximation yields an effective
time-independent description, in which the qubit frequencies are
replaced by their detunings from the drive,
$\delta\omega_i=\omega_i-\omega_{\rm dr}$.

For a resonantly driven single-qubit junction, the entropy conductance
follows directly from the general expression derived above and can be
decomposed into incoherent and coherent contributions\cite{renyi_heatEngine},
\begin{align}
\mathcal{I}^{\rm 1q}_{\rm coh}
&=
\Gamma_p
\left|\rho_{01}\right|^{2},
\label{eq:I_1q_coh}\\
\mathcal{I}^{\rm 1q}_{\rm inc}
&=
\Gamma_p^{\downarrow}\rho_{11}
-
\Gamma_p^{\uparrow}\rho_{00}.
\label{eq:I_1q_inc}
\end{align}
The incoherent contribution depends exclusively on level populations and
therefore reflects entropy exchange associated with stochastic
transitions. In contrast, the coherent contribution is proportional to
the magnitude of the off-diagonal density-matrix element and originates
from quantum coherence generated by the external drive.

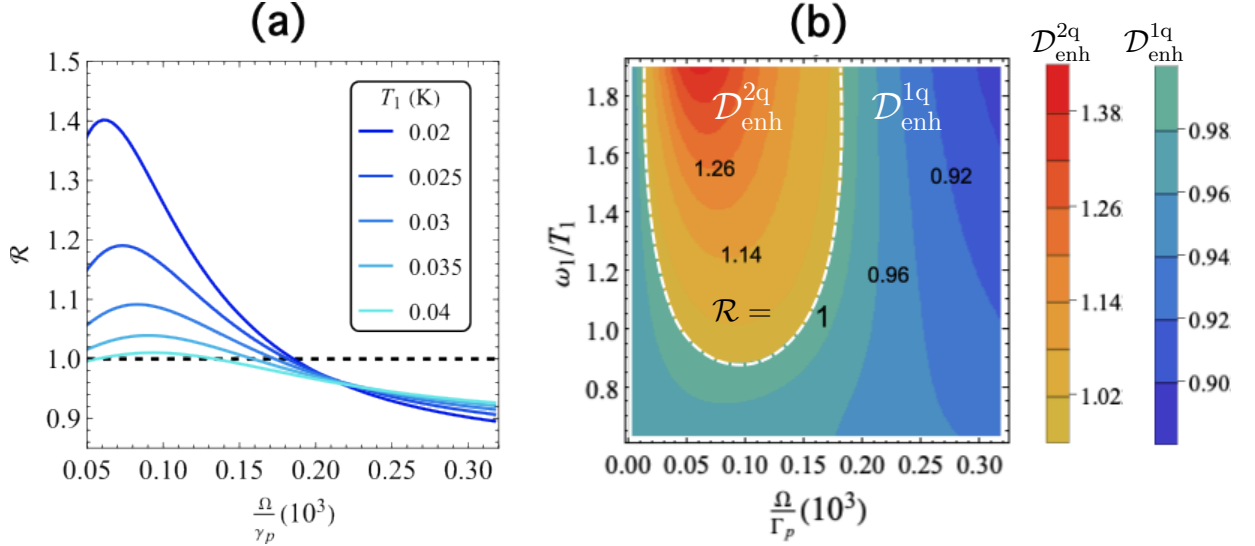
\begin{figure*}[ht]
    \includegraphics[width=0.9\linewidth]{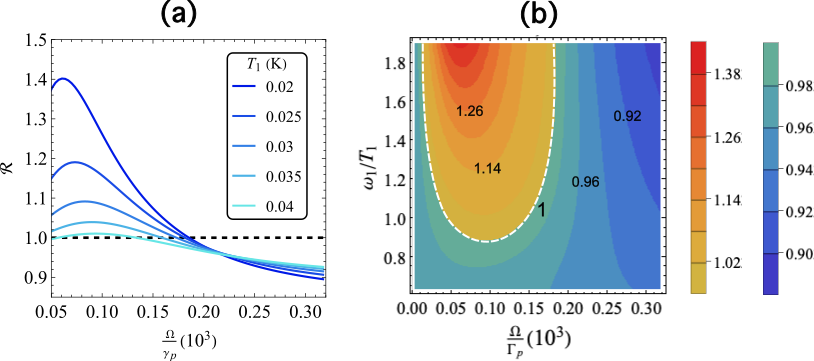}
    \put(-195,84){\color{black} \large {$\mathcal{R}=$}}
    \put(-195,160){\color{white} \Large $\mathcal{D}_{\rm enh}^{\rm 2q}$}
    \put(-135,160){\color{white} \Large $\mathcal{D}_{\rm enh}^{\rm 1q}$}
    \put(-75,185){\color{black} \large $\mathcal{D}_{\rm enh}^{\rm 2q}$}
    \put(-40,185){\color{black} \large $\mathcal{D}_{\rm enh}^{\rm 1q}$}
    \caption{ \textbf{Entropy-conduction enhancement factor ($\mathcal{R}$) at the stationary state as a function of $\Omega$ and $T_1$.} We consider the main reservoir temperature $T_1$ to be between 20-60 mK while probe reservoir's temperature is nearly zero. (\textbf{a}) In colder environment the two-qubit can be programmed to be faster and reach a peak value at weak driving regime ($\Omega \ll \Delta$). Increasing the driving power flips the ratio to make single-qubit faster than two qubit junction. 
    (\textbf{b}) Contour map of $\mathcal{R}$ in the $(\Omega,T_1)$ parameter plane. The dashed contour marks $\mathcal{R}=1$, separating the regime
    of two-qubit flow dominance $\mathcal{R}>1$
    from the opposite regime $\mathcal{R}<1$, where the single-qubit junction yields the
    larger response. Here in the two-qubit junction $J/\Delta=0.6$ where, $\Delta=(\omega_1-\omega_2)$. 
    }
    \label{fig:f_ratio}
\end{figure*}
Applying the same formalism to the driven two-qubit junction yields
\begin{align}
\mathcal{I}^{\rm 2q}_{\rm coh}
&=
\Gamma_p
\left|
\rho_{01,00}
+
\rho_{11,10}
\right|^{2},
\label{eq:I_2q_coh}\\
\mathcal{I}^{\rm 2q}_{\rm inc}
&=
\Gamma_p^{\uparrow}
\left(
\rho_{00,00}
+
\rho_{10,10}
\right)
-
\Gamma_p^{\downarrow}
\left(
\rho_{01,01}
+
\rho_{11,11}
\right).
\label{eq:I_2q_inc}
\end{align}
Here $\rho_{mn,lk}=\bra{mn}\rho_S\ket{lk}$ denotes an element of the
two-qubit reduced density matrix. The incoherent contribution depends
only on the diagonal elements of $\rho_S$ in the bare basis, whereas
the coherent contribution depends on off-diagonal coherences associated
with transitions of the probe-coupled qubit. Unlike in the single-qubit
case, the coherent contribution in the two-qubit junction involves a
collective superposition of multiple transition pathways. Entropy
transport therefore becomes sensitive not only to local coherences but
also to the multi-qubit structure of the quantum state.

This result shows that enlarging the Hilbert space qualitatively modifies
the mechanisms governing entropy exchange. Quantum correlations and
drive-induced coherences generated by multi-qubit operations can
therefore leave measurable signatures in the entropy conductance.
\subsection{Comparison : two-qubit vs single-qubit}
Our objective is to determine how the internal structure of the quantum
junction influences the entropy conductance into a probe reservoir. As we
show below, entropy transport is not governed solely by reservoir
temperatures or coupling strengths. Instead, it is strongly shaped by the
 the number of qubits and their junction parameters in the quantum circuit mediating the transport process. The comparison between the single- and
two-qubit architectures therefore demonstrates that entropy flow can
carry direct signatures of the underlying quantum operation implemented
within the junction.

To quantify how circuit complexity modifies entropy transport, we introduce 
the \emph{enhancement} factor $\mathcal{R}\equiv 
    \mathcal{I}^{\rm 2q}/\mathcal{I}^{\rm 1q}$, which compares the stationary state von Neumann entropy conductance (\ref{eq:von}) of the driven
two-qubit junction (\ref{eq:I_2q_coh},\ref{eq:I_2q_inc}) with that of the driven single-qubit junction (\ref{eq:I_1q_coh},\ref{eq:I_1q_inc}).

Values $\mathcal{R}>1$ indicate that the two-qubit junction enhances entropy
conductance relative to the single-qubit junction. This enhancement originates
from the additional quantum degrees of freedom and conditional dynamics
available in the two-qubit architecture. Conversely, $\mathcal{R}<1$ indicates
a suppression of entropy flow relative to the single-qubit reference.
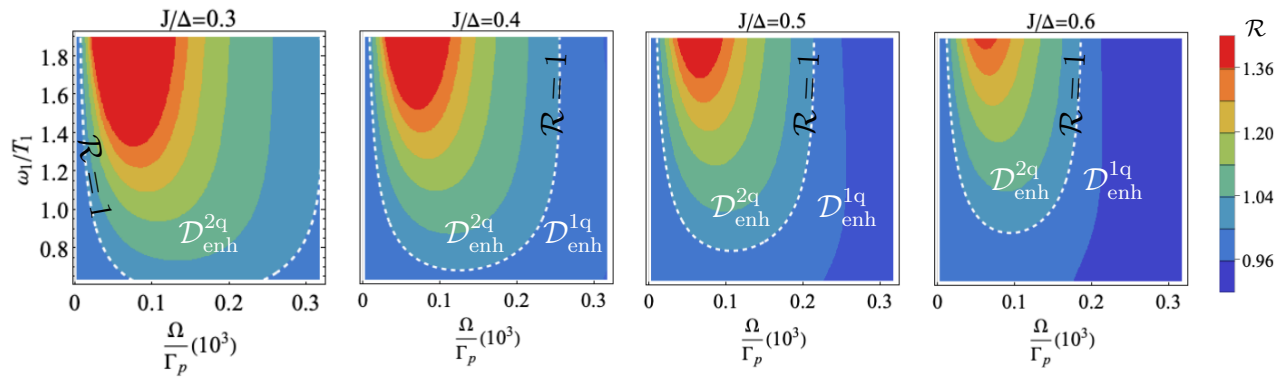
\begin{figure*}[ht]
    \centering
    \includegraphics[width=0.95\linewidth]{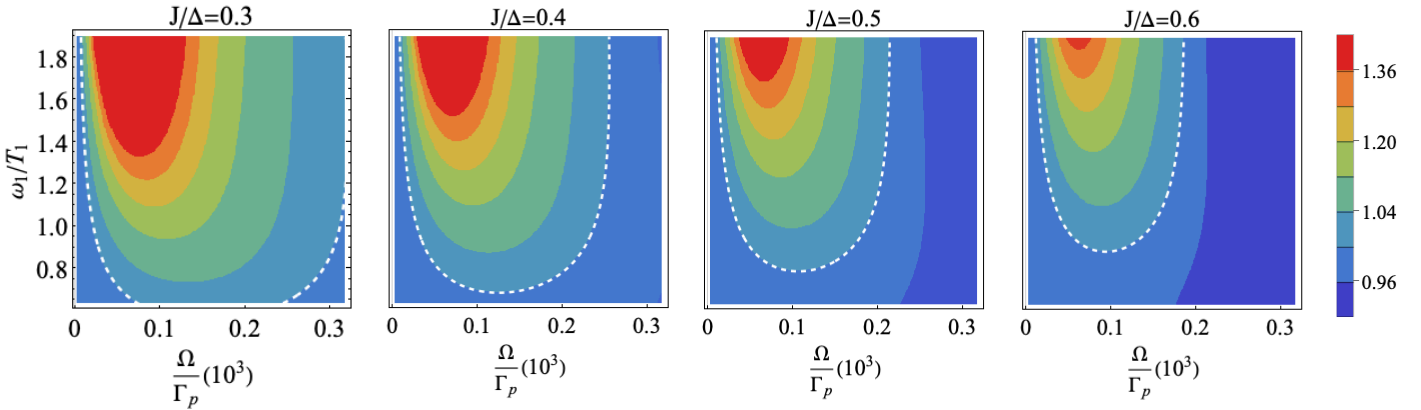}
    \put(-458,90){\color{black} \large \rotatebox{-80}{$\mathcal{R}=1$}}
    \put(-420,50){\color{white} \large $\mathcal{D}_{\rm enh}^{\rm 2q}$}
    \put(-284,90){\color{black} \large \rotatebox{87}{$\mathcal{R}=1$}}
    \put(-320,50){\color{white} \large $\mathcal{D}_{\rm enh}^{\rm 2q}$}
   \put(-283,50){\color{white} \large $\mathcal{D}_{\rm enh}^{\rm 1q}$} 
   \put(-189,90){\color{black} \large \rotatebox{87}{$\mathcal{R}=1$}}
   \put(-220,60){\color{white} \large $\mathcal{D}_{\rm enh}^{\rm 2q}$}
   \put(-180,60){\color{white} \large $\mathcal{D}_{\rm enh}^{\rm 1q}$} 
   \put(-89,90){\color{black} \large \rotatebox{87}{$\mathcal{R}=1$}}
    \put(-115,70){\color{white} \large $\mathcal{D}_{\rm enh}^{\rm 2q}$}
    \put(-80,70){\color{white} \large $\mathcal{D}_{\rm enh}^{\rm 1q}$}
      \put(-19,128){\color{black} $\mathcal{R}$}
    \caption{
   \textbf{Entropy-conduction enhancement factor ($\mathcal{R}$) at the stationary state for varying $J/\Delta$.}
    Enhancement factor $\mathcal{R}$ for four representative values of the
    dimensionless coupling ratio $J/\Delta=0.3,0.4,0.5$ and $0.6$.
    Regions with $\mathcal{R}>1$ identify regimes in which the driven
    two-qubit junction transports entropy more efficiently than the driven
    single-qubit junction. The enhancement region expands as $J/\Delta$ is
    reduced, showing that entropy transport can be tuned by controlling the
    hybridization between the two qubits.}
    \label{fig:f_ratioJ}
\end{figure*}

Fig. \ref{fig:f_ratio}\blu{a} shows the stationary entropy conductance enhancement factor $\mathcal{R}$ for discrete temperatures $T_1$ and continuously varying  drive amplitude $\Omega$. The two-qubit junction exhibits a larger stationary entropy conductance in the weak-driving regime, $\frac{\Omega}{\Delta}$<0.01. In this regime, the enhancement can reach above $40\%$ at lower temperatures. As the drive amplitude is increased, however, the ratio crosses
over and the single-qubit junction becomes the more efficient entropy
conductor. 

In general, the conductance depends on the microscopic parameters of the junction. In particular, for two-qubit junction
$\mathcal{I}^{\rm 2q}
=
\mathcal{I}^{\rm 2q}
(\Omega,J,T_1,\omega_1,\omega_2,\omega_{\rm dr})$, 
whereas for the single-qubit junction
$\mathcal{I}^{\rm 1q}
=
\mathcal{I}^{\rm 1q}
(\Omega,T_1,\omega_{0},\omega_{\rm dr})$.
Fig.~\ref{fig:f_ratio}\blu{b} shows a density plot for the enhancement factor for a representative
choice of parameters, where $J/\Delta$ is fixed with  $\Delta=\omega_1-\omega_2$. The dashed white curve corresponds to the condition
$\mathcal{I}^{\rm 2q}(\Omega,T_1)
=
\mathcal{I}^{\rm 1q}(\Omega,T_1)$, 
or equivalently $\mathcal{R}=1$. Along this boundary the two junctions are indistinguishable with respect to entropy conductance. Inside this boundary, a finite region emerges, in which the two-qubit
junction transports entropy more efficiently than the single-qubit junction.
We denote this enhancement domain by
\begin{equation}
\mathcal{D}_{\rm enh}^{\rm 2q}
=
\left\{
(\Omega,T_1)
\,\middle|\,
\mathcal{R}(\Omega,T_1)>1
\right\},
\end{equation}
corresponding to the red region in Fig.~\ref{fig:f_ratio}\blu{b}, while the complimentary region $\mathcal{D}_{\rm enh}^{\rm 1q}$ is indicated in in blue spectrum. 

The observed enhancement, results from a competition between three
processes: drive-induced excitation, inter-qubit hybridization, and
reservoir-induced relaxation. The balance of these mechanisms determines
whether quantum dynamics amplify or suppress entropy transport.

We now distinguish two regimes separated by the drive amplitude $\Omega_0$ for which $\omega_1/T_1$ reaches a minima on $R=1$ (dashed white curve) in Fig. \ref{fig:f_ratio}\blu{b}.
First, consider the strong-driving regime, $\Omega>\Omega_0$. In the
two-qubit junction, the drive-induced transition rate is proportional to
$\Omega\cos\theta$, with $\theta
    =
    \frac{1}{2}
    \arctan\left(\frac{2J}{\Delta}\right)$, whereas for resonantly driven single-qubit
case the corresponding rate is proportional to $\Omega$ as discussed in Appendix~\ref{app:gates}. Since
$\Omega\cos\theta \leq \Omega$, the drive induced excitation in
two-qubit junction is less than the single-qubit junction. Therefore,
when the drive is a dominant source of injected energy, the single-qubit
junction has an advantage. This explains why the two-qubit enhancement
region, $\mathcal{D}_{\rm enh}^{\rm 2q}$, shrinks as $\Omega$ is increased.

We next consider the weak-driving regime, $\Omega<\Omega_0$. In the
single-qubit junction, energy supplied by the hot reservoir can be transported
to the probe reservoir directly via $q_1$. In contrast, in the two-qubit junction, this transport requires that excitation first be transferred from $q_1$ to $q_2$. However, this transfer is limited by the inter-qubit coupling J and hence the two-qubit system is at disadvantage. Nevertheless, when energy transport is also activated by the drive this scenario can be flipped. But the drive is inefficient at very small $\Omega$ to make $\mathcal{R}>1$. 
The drive-assisted transport becomes more effective as the drive amplitude $\Omega$ increases in this region, and the
enhancement region $\mathcal{D}_{\rm enh}^{\rm 2q}$ expands over a broader range of reservoir temperatures $T_1$.  In the limit of vanishingly weak drive, $\Omega \ll \Omega_0$ the two-qubit junction is
disadvantaged and shows suppressed entropy conductance.

Finally, Fig.~\ref{fig:f_ratioJ} shows that the enhancement region can also be
controlled by tuning the hybridization parameter $J/\Delta$. Reducing
$J/\Delta$ increases $\cos\theta$ and therefore strengthens the effective
drive-induced transition rate $\Omega\cos\theta$. As a result, the domain
$\mathcal{D}_{\rm enh}^{\rm 2q}$ extends over a wider region of parameter
space. This confirms that entropy-conductance provides a thermodynamic signature of multi-qubit
information processing and offers a direct means of distinguishing
single-qubit and multi-qubit transport mechanisms in driven quantum circuits. 

\section{Discussion}\label{sec:discussion}

Beyond the quantitative comparison presented above, the results reveal several non-trivial features of entropy transport in driven quantum junctions. These features provide physical insight into why the two-qubit architecture can outperform the single-qubit junction over a finite range of driving amplitudes, despite having a nearly asymptotic behaviour at strong drive. 

\begin{figure}[H]
     \centering
        \includegraphics[width=0.5\textwidth]{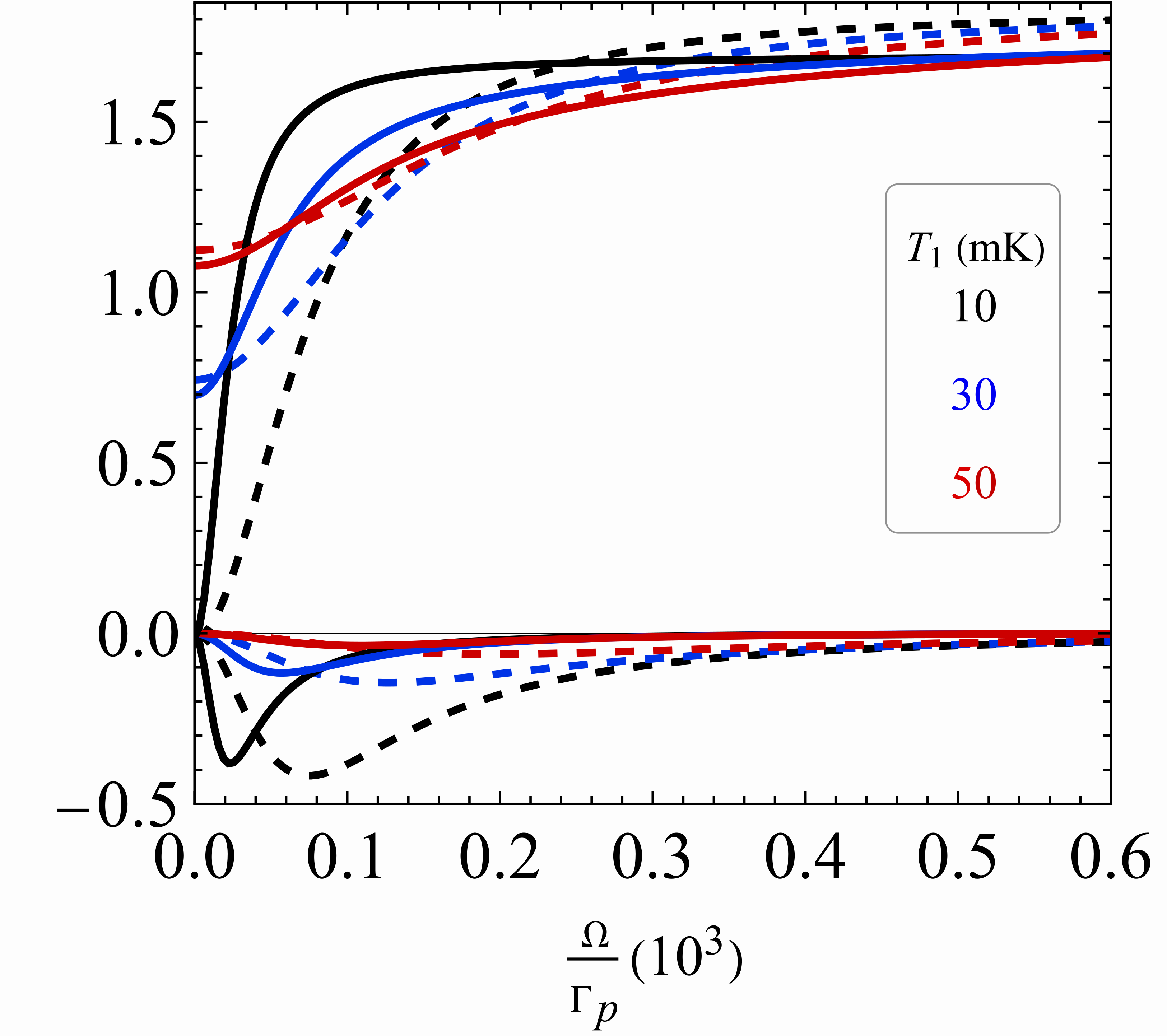}
        \put(-140,190){\color{black} \large $\mathcal{I}_{inc}/\Gamma_p$}
        \put(-120,70){\color{black} \large $\mathcal{I}_{coh}/\Gamma_p$}
        \put(-250,100){\color{black} \large \rotatebox{90}{Entropy flow ($\mathcal{I}$)}}
        \caption{
        \textbf{Coherent and incoherent contributions to entropy flow.}
        The total entropy flow $\mathcal{I}$ is decomposed into incoherent and coherent components, $\mathcal{I}_{\rm inc}$ and $\mathcal{I}_{\rm coh}$, respectively. The incoherent contribution is positive, whereas the coherent contribution is negative. The probe temperature is fixed at $T_p=10\,{\rm mK}$, while the temperature $T_1$ of reservoir $R_1$ is varied. Entropy flow in two-qubit junction $\mathcal{I}^{\rm 2q}$ (solid lines), converges at weaker driving compared to the the entropy flow in single-qubit junction $\mathcal{I}^{\rm 1q}$ (dashed lines).
        }
        \label{fig:f_entropySeparated}
\end{figure}

The stationary state of the quantum junction is obtained from the Lindblad dynamics of the driven open quantum system. Once this non-equilibrium stationary state is reached, the entropy flow into the probe reservoir follows from Eq.~(\ref{eq:von}). The resulting entropy current depends sensitively on the driving amplitude $\Omega$, and this dependence is further modulated by the thermal imbalance between the reservoirs. To quantify this behaviour, Fig.~\ref{fig:f_entropySeparated} shows the entropy conductance for single-qubit (dashed lines) and two-qubit (solid lines) systems as a function of driving amplitude $\Omega$ for several values of the reservoir temperature $T_1$, while keeping the probe temperature fixed at $T_p \approx 0$. In particular, the separation of the entropy flow into incoherent $\Iincoh$ and coherent $\Icoh$ contributions clarifies how stationary non-equilibrium transport and drive-induced quantum coherence compete in determining the total entropy current.

A central observation is that the entropy flow saturates at large driving amplitudes in both architectures. However, the approach to saturation is markedly different. In the two-qubit junction, the entropy flow rises more steeply with $\Omega$ and reaches its saturation regime at a smaller drive amplitude than in the single-qubit case. This faster convergence is responsible for the finite parameter region in which the enhancement ratio satisfies $\mathcal{R}>1$. Physically, this behaviour reflects the fact that the driven two-qubit junction does not merely act as a larger dissipative system. Instead, the driving interaction opens a conditional transport channel in which the state of one qubit (the driven one) controls the excitation dynamics of the other (the undriven one). The efficiency of this additional pathway determines the range of $\Omega$ for which enhancement occurs.
The enhancement is nevertheless bounded. At sufficiently large $\Omega$, the entropy flow in both architectures becomes dominated by the drive-induced energy exchange with the junction. In this regime, the single-qubit junction has a slight asymptotic advantage, and the entropy flow in the single-qubit setup saturates at a marginally larger value. 
As a result, the region of $\mathcal{D}^{2q}_{\rm enh}$ remains finite rather than extending indefinitely to arbitrarily strong drive. This shows that the advantage of the two-qubit junction is not a trivial consequence of stronger driving, but instead arises from an intermediate regime where coherent two-qubit dynamics and the reservoir-induced transport are optimally balanced. 

This interpretation is further supported by the decomposition into coherent and incoherent contributions. In the absence of driving, $\Omega=0$, a quantum junction connecting reservoirs of equal temperature relaxes to the thermal stationary state $ \rho_S^{\rm eq}=\exp(-\beta_0 H_S)/Z$, where $\beta_0=(k_{\rm B}T_0)^{-1}$ with reservoirs are all held at the same temperature $T_0$. In this equilibrium limit, detailed balance is restored---the net entropy flow vanishes, as seen in Fig.~\ref{fig:f_entropySeparated} for the curve with $T_1=T_p$ at $\Omega=0$.

By contrast, when the two reservoirs are maintained at different temperatures, the junction relaxes to a non-equilibrium stationary state even in the absence of coherent driving. In this regime, a finite entropy current is sustained between the reservoirs and is mediated solely by the quantum junction. Crucially, this undriven component originates entirely from the single-world, or incoherent contribution $\mathcal{I}_{\rm inc}$, defined in Sec.~\ref{app:entropy}. Consistently, Fig.~\ref{fig:f_entropySeparated} shows that at $\Omega=0$ the coherent component vanishes, $\mathcal{I}_{\rm coh}=0$, for all temperature biases, whereas the incoherent contribution remains finite whenever $T_1\neq T_p$. This phenomenology parallels the entropy transport previously identified in two-level single-qubit junctions~\cite{renyi_heatEngine} and in three-level photovoltaic cell ~\cite{Ansari_2017e}.

\begin{figure*}[ht]
    \centering
    \includegraphics[width=0.9\linewidth]{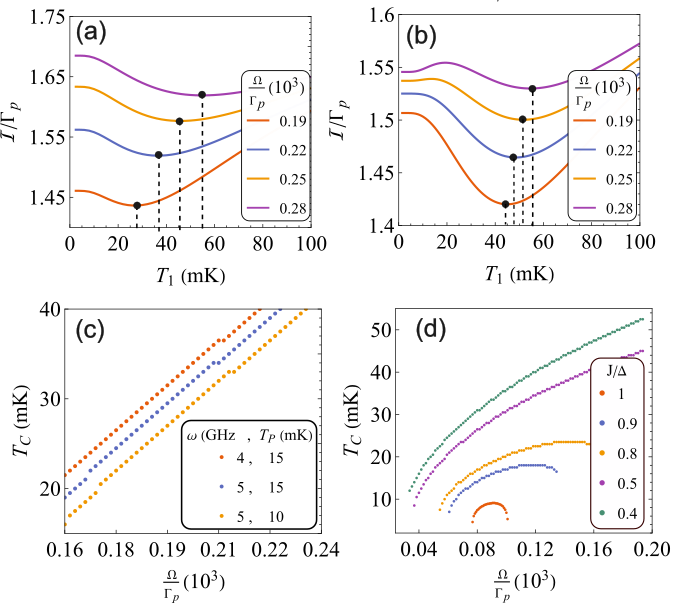} \vspace{-0.1in}
    \caption{
\textbf{Critical temperatures for the onset of negative differential entropy conductance.}
The upper panels show the stationary dimensionless entropy flow, $\mathcal{I}/\Gamma_p$, as a function of reservoir temperature $T_1$ for several values of the drive amplitude $\Omega$ in
\textbf{(a)} a driven single-qubit junction and
\textbf{(b)} a driven two-qubit junction with $J/\Delta=0.5$.
Black markers indicate the minima of the entropy-flow curves and their corresponding temperatures are the onset temperatures $T_C$, at which the entropy current is most strongly suppressed.
The lower panels show the resulting dependence of $T_C$ on $\Omega$ for
\textbf{(c)} the single-qubit junction and
\textbf{(d)} the two-qubit junction.
}
\label{fig:criticalTemp}
\end{figure*}
The coherent contribution has a qualitatively different physical origin. It is not a perturbative correction to ordinary thermal transport, but rather a genuinely drive-induced channel of entropy flow. For $\Omega\neq0$, the drive activates the Cross-World, or coherent contribution introduced in Appendix \ref{app:entropy}. The negative sign of $\mathcal{I}_{\rm coh}$, as seen in Fig.~\ref{fig:f_entropySeparated}, indicates that the coherent processes partially oppose the incoherent entropy current. This counterflow is visible for all driven curves. Quantitatively, near its maximum, the coherent contribution can reach a magnitude that is half of the incoherent contribution, demonstrating that it is not a negligible contribution. We therefore retain both coherent and incoherent contributions throughout the calculation of the total entropy conductance. 
\subsection*{Negative differential entropy conductance}
A striking consequence of entropy transport through a coherently driven quantum junction is that the stationary entropy current need not grow monotonically with the imposed thermal bias. As shown in Figs.~\ref{fig:criticalTemp}\blu{a} and \ref{fig:criticalTemp}\blu{b}, the dimensionless entropy flow,  $\mathcal{I}/\Gamma_p$, develops a pronounced minimum as the temperature $T_1$ of the main reservoir is increased.  Equivalently, the regions on the left of these minimas in which increasing $T_1$ at fixed drive amplitude $\Omega$ reduces the entropy flow into the probe, defines a regime of negative differential entropy conductance (NDEC),
\begin{equation}
    \frac{\partial \mathcal{I}}{\partial T_1}<0 ,
\end{equation}
the entropy-current analogue of negative differential thermal conductance(NDTC).

The single- and two-qubit junctions realize this mechanism in qualitatively different ways. For the single-qubit junction, NDEC appears robustly across the explored parameter window. In the two-qubit junction, the effect is controlled by the ratio between inter-qubit coupling and detuning. In the detuning-dominated regime, $J/\Delta<1$, the dressed transition structure remains sufficiently resolved to support a non-monotonic entropy response. By contrast, for $J/\Delta>1$, stronger hybridization reorganizes the effective transition channels and suppresses NDEC within the explored range. Thus, NDEC in the two-qubit architecture is not simply induced by driving; it is selected by the internal spectral structure of the junction.

To quantify this effect, we define $T_C$ as the temperature at which $\mathcal{I}$ reaches its minimum. The black markers in Figs.~\ref{fig:criticalTemp}\blu{a}, \ref{fig:criticalTemp}\blu{b} identify these minima for different drive amplitudes. Figs. \ref{fig:criticalTemp}\blu{c},\ref{fig:criticalTemp}\blu{d} show the resulting dependence of $T_C$ on $\Omega$: in Fig.~\ref{fig:criticalTemp}\blu{c} for representative single-qubit parameters, specified by $\omega_0$ and $T_p$, and in Fig.\ref{fig:criticalTemp}\blu{d} for different values of $J/\Delta$ in the two-qubit junction. The effect is suppressed as we increase $J/\Delta$ upto 1 and completely vanishes after that. The comparison shows that the two-qubit junction can enter the same entropy-blockade regime at substantially weaker drive amplitudes than the single-qubit junction, indicating a more power-efficient route to suppressing entropy injection into the probe.

Negative differential entropy conductance therefore represents more than a transport anomaly. It identifies a regime in which entropy flow is governed by the drive-dressed spectrum of the quantum junction rather than by the temperature gradient alone. In programmable circuits, tuning $\Omega$, $J$ and $\Delta$ provides a gate-level thermodynamic control knob, enabling a switch between monotonic entropy transport and a regime in which increasing thermal bias suppresses the entropy current. This establishes a multi-qubit mechanism for engineering non-equilibrium entropy flow and suggests operating regimes relevant for reservoir protection and quantum refrigeration.
\subsection*{Resonances}\label{sec:peaks}
To maximize the effect of flow of entropy, motivated by the cross-resonance gate in superconducting quantum computing \cite{CR_exp,xu2024lattice,xu2025parity-050}, we examined various driving frequencies at which we drive the control qubit in the two-qubit junction, while keeping all other parameters fixed.  This helps to determine the optimal driving frequency $\wdr$,  at which stationary entropy flow $\mathcal{I}$ is amplified.

\begin{figure}[H]
    \centering
    \includegraphics[width=0.49\textwidth]{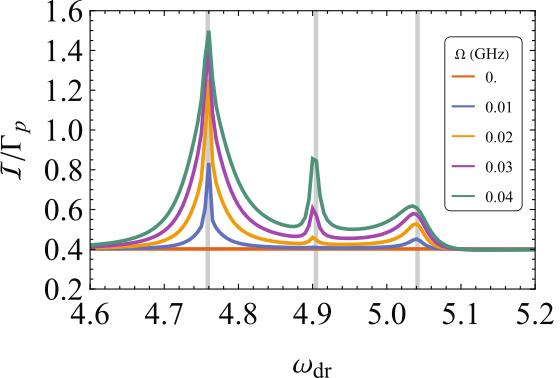}
    \put(-165,153){\color{black} \large {(A)}}
    \put(-118,107){\color{black} \large {(B)}}
    \put(-75,84){\color{black} \large {(C)}}
    \caption{
    \textbf{Drive-frequency dependence of the entropy flow.}
    Stationary entropy flow $\mathcal{I}$ as a function of the drive frequency $\omega_{\rm dr}$.
    For the two-qubit junction, three entropy-flow maxima appear at the three anticrossings in the energy spectrum of $H_S$:
    \textbf{(A)} $\omega_{\rm dr}\simeq \min\{E_4-E_3\}$,
    \textbf{(B)} $\omega_{\rm dr}\simeq \min\{E_3-E_2\}$, and
    \textbf{(C)} $\omega_{\rm dr}\simeq \min\{E_2-E_1\}$.
    }
    \label{fig:f_peaks}
\end{figure}

For the single-qubit junction, the entropy flow exhibits a single maximum at resonance, $\omega_{\rm dr}=\omega_0$, as expected for coherent resonant driving. By contrast, the two-qubit junction displays three distinct maxima, as shown in Fig.~\ref{fig:f_peaks}. In the main results, we choose $\omega_{\rm dr}$ to coincide with the largest entropy-flow in each architecture. For the two-qubit junction, this maximum corresponds to driving the higher-frequency qubit near the transition frequency of the lower-frequency qubit, which is the operating condition associated with the leftmost peak and yields the largest entropy current.
The remaining two peaks reveal additional entropy-transport channels. The rightmost peak corresponds predominantly to driving the qubit near its own transition frequency, producing a weaker and slower entropy transfer to the probe reservoir. The middle peak is more subtle: it arises from resonant activation of a dressed transition near the mean frequency of the two qubits, where an avoided crossing appears between the one-excitation states, $\ket{01}$ and $\ket{10}$ as shown in Fig.\ref{fig:f_eig_evol0} in Appendix \ref{app:anticrossings}. This resonance has no direct analogue in the single-qubit junction. At this intermediate resonance, the associated population dynamics is distinct, but the distinction is not as visually prominent as compared to that at the other two resonances, as discussed in Appendix~\ref{app:anticrossings}. However, a significant peak still appears at this point,  indicating a considerable involvement of reservoir interactions in building this effect.
\section{Conclusion}\label{sec:conclusion}
We have shown that entropy transport through driven quantum junctions can be engineered by the structure of the junction itself. Using the Keldysh-diagramatic approach based purturbative method, we evaluated the von Neumann entropy flow into a probe reservoir for a general setting as well as the example of a driven two-qubit junction and the result was compared with a driven single-qubit junction. This information-theoretic formulation treats the entropy current as the rate of change of the probe-reservoir entropy, rather than as a quantity inferred indirectly from heat or energy transport. It therefore provides a direct route to characterizing entropy flow in open quantum circuits.

The comparison between the two architectures reveals that entropy conductance is not determined by the reservoir bias alone. It is strongly shaped by the internal quantum dynamics of the junction, by the drive frequency and amplitude, and by the dressed transition structure through which entropy is transferred. In the single-qubit case, entropy transport is governed by a single resonant channel. In the two-qubit case, additional dressed-state pathways appear, allowing entropy flow to be redistributed, enhanced or suppressed. The two-qubit junction can therefore outperform the single-qubit junction in selected parameter regimes, achieving comparable or larger entropy conductance at lower driving strength.

The underlying mechanism is revealed by some additional results including the emergence of coherent contributions to the entropy current and of negative differential entropy conductance. In this regime, increasing the temperature bias can reduce, rather than increase, the entropy flow into the probe reservoir. This counter-intuitive response identifies a regime in which entropy transport is controlled by the spectral and coherent properties of the junction and not simply by the thermal gradient.

Our results lay the groundwork for using multi-qubit gate physics not only as computational primitives, but also to engineer entropy flow in driven non-equilibrium quantum devices. By tuning gate-level parameters such as drive amplitude ($\Omega$), drive frequency ($\omega_{dr}$), inter-qubit coupling ($J$) and detuning ($\Delta$), one can shape the entropy current, switch between monotonic and negative-differential transport, and design regimes of reservoir protection or refrigeration. This establishes a route towards quantum circuits in which information processing and entropy management are controlled within the same hardware platform.

Several directions follow naturally from this work. Extending the analysis to larger multi-qubit networks would clarify how entropy transport scales with circuit complexity and connectivity. Time-dependent protocols beyond the stationary regime could reveal how entropy can be dynamically routed, stored or suppressed during finite-time gate operations. Finally, combining entropy-flow diagnostics with realistic noise models and experimentally accessible observables may provide a practical framework for designing programmable quantum thermal machines based on superconducting and other solid-state quantum processors.
\section*{Acknowledgments}
We thank Julian Rapp and Alwin van Steensel for fruitful discussions. 

\bibliography{entropy_ref, references2, MAbib}
\onecolumngrid
    \section*{Appendix}
    \appendix
\section{Entropy flow expression: derivation with Keldysh diagramatic approach}\label{app:entropy}
\begin{figure*}
    \centering
    \includegraphics[width=0.9\linewidth]{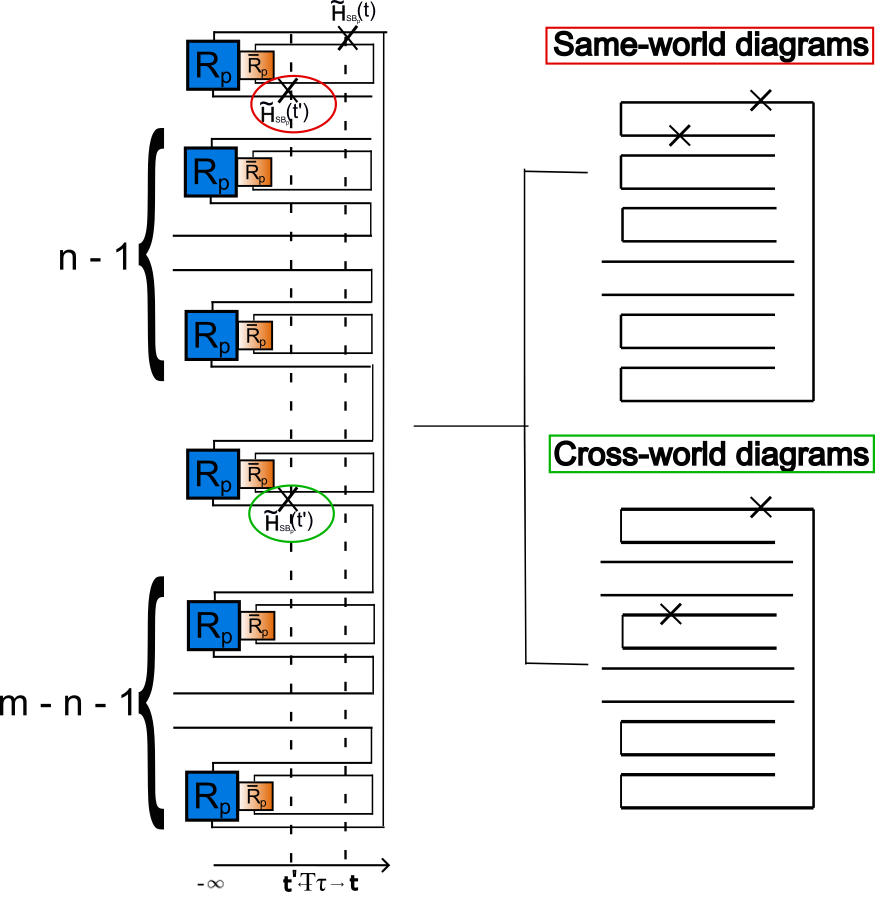}
    \caption{Keldysh based diagrammatic approach : an example term with m-copies of $\rho(\tilde{t})$ is shown in order to  calculate R\'enyi entropy flows $\mathcal{I}_{m}^{R_p}$ to the probe reservoir $R_p$(in blue). Rest of the degrees of freedom $\bar{R}_p$(in orange), are traced out (indicated by closed contours) within each copy. Interactions $\tilde{H}_{SB_p}(t)$ and $\tilde{H}_{SB_p}(t')$ are given by cross $\cross$. Two interactions lying in the same copy correspond to Single-world diagrams. Two interactions lying in different copies correspond to Cross-world diagrams.}
    \label{fig:keldysh}
\end{figure*}
This section outlines the procedure for obtaining the expression for entropy flows used in Sec. \ref{sec:entropyFlow}. We apply the derivation explained in \cite{entropy} to our model.
The complete Hamiltonian is given in Eq.(\ref{eq:Htot}). The desired quantity is the R\'enyi entropy flow, $\mathcal{I}_{m}^{R_p}=-\frac{d}{dt}ln\,\left(Tr\left[\rho_{R_p}^m\right]\right)$ to the probe reservoir $R_p$. Let us expand this to get the entropy flow in the form
\begin{align}\label{eq:expand_entropy}
  \mathcal{I}_{m}^{R_p}=&=-ln\,\left(Tr_{R_2}\left[\dot{\rho}_{R_p}\rho_{R_p}^{m-1}+\rho_{R_2}\dot{\rho}_{R_p}\rho_{R_p}^{m-2}\right.\right. \nonumber\\
    & \qquad\qquad \left. \left. +\dots+\rho_{R_p}^{m-1}\dot{\rho}_{R_p}\right]\right).
\end{align}

Time derivative of $\rho_{R_p}$ is denoted by a dot on top, $\dot{\rho}_{R_p}$. We consider the local approach \cite{bridging_local_global,local_global}, meaning that the interaction of $R_p$ with the quantum system has no direct effect from the drive, coupling between the qubits and other reservoirs. The effect of these elements arises only through the system state evolution. As a result, $\rho_{R_p}$ and $\dot{\rho}_{R_p}$ depend only on the interaction term involving the probe reservoir $R_p$ in the interaction picture. In our model this term is 
$$\tilde{H}_{SB_p}=\,e^{-i\omega_{p}\,t}\,\hat{\sigma}_p^-\text{O}_{B_p}(t)+\,e^{+i\omega_{p}\,t}\hat{\sigma}_p^+\text{O}^\dagger_{B_p}(t)$$
where $\text{O}_{B_p}(t)=\sum_k\,e^{-i\omega_k\,t}\,b^-_{p,k}\hat{b}_{p,k}\,+\,e^{+i\omega_k\,t}b^+_{p,k}\,\hat{b}^\dagger_{p,k}$ and subscript $p$ denotes the operator or frequency of the qubit to which the probe reservoir $R_p$ is coupled. Operators are either denoted by a Capital letter e.g. $\text{O}$ or with a hat on top e.g. $\hat{o}$. Truncating the expressions to second order we have
$$ \rho_{R_p}=R^{(0)}+R^{(1)}+\mathcal{O}(2)$$
where
\begin{align*}
  R^{(0)}&=\rho_{R_p}(-\infty) \\ 
  R^{(1)}&=Tr_{\bar{R}_p}\int_{-\infty}^t\,dt'\,\left(\frac{1}{i\hbar}\right)\left[\tilde{H}_{SB_p}(t'),\rho(-\infty)\right]
\end{align*}
and $$ \dot{\rho}_{R_p}=\Delta^{(1)}+\Delta^{(2)}+\mathcal{O}(3)$$
where
\begin{align*} 
 \Delta^{(1)} &=Tr_{\bar{R}_p}\,\left(\frac{1}{i\hbar}\right)\left[\tilde{H}_{SB_p}(t'),\rho(-\infty)\right]\\
   \Delta^{(2)} & =Tr_{\bar{R}_p}\int_{-\infty}^t\,dt'\,\left(\frac{1}{i\hbar}\right)^2\\
    &\left[\tilde{H}_{SB_p}(t),\left[\tilde{H}_{SB_p}(t'),\rho(-\infty)\right]\right].
\end{align*}
where $\bar{R}_p$ denotes complement of $R_p$ i.e. $q_1+R_1$ in one qubit  setup and $q_1+q_2+R_1$ in two qubit  setup. Only second order terms give non-zero contribution. Each term in Eq.(\ref{eq:expand_entropy}) contains $\dot{\rho}_{R_p}$, which has lowest order one. Therefore, we have not considered the second order term in ${\rho}_{R_2}$, since it is never used in the second order perturbation.
    
We use Keldysh contours to keep track of all the terms and in turn it makes our calculations easier. Consider m-copies of the density matrix of the whole system $\rho(\tilde{t})$ as shown in Fig. \ref{fig:keldysh} where time $\tilde{t}$ runs from $-\infty$ to $t$. Ket part $\ket{}$ goes from left to right and Bra part $\bra{}$ goes from right to left on the time line given at the bottom. The trace over $\bar{R}_p$ is shown by closed contour within each copy. Each interaction $\tilde{H}_{SB_p}$ either at time $t$ or at $t'$ is shown by a cross. The diagrams get divided into two categories, the ones containing $\Delta^{(2)}$ and $\Delta^{(1)}R^{(1)}$. For terms with $\Delta^{(2)}$, both interactions take place in the same copy. Whereas, for terms involving $\Delta^{(1)}R^{(1)}$, two interactions happen in two different copies. Hence the names Same-world(SW) and Cross-world(CW) respectively.

An example, showing this difference is given in Fig. \ref{fig:keldysh}. The first interaction term $\tilde{H}_{SB_p}(t)$ is chosen to lie on the first copy, out of m possible choices. It lies on the upper half of the first copy, which again is a choice made out of the two terms in the commutator. If the second interaction term $\tilde{H}_{SB_p}(t')$ lies in the same copy as $\tilde{H}_{SB_p}(t)$, as circled in red, then we have an example of a Same-world diagram. If it lies on a differnt world, as circled in green, then we have a Cross-world diagram. In Cross-world diagrams we have remaining $(m-1)$ copies to choose from. Once the copy for $\tilde{H}_{SB_p}(t')$ is chosen, we again have a choice between two terms in the commutator containing $\tilde{H}_{SB_p}(t')$. In total, there are $4\,m$ SW diagrams and $4m(m-1)$ CW diagrams. When all these contributons are represented as diagrams as shown in Fig.\ref{fig:keldysh}, we see that most of the diagrams cancel out. This is because all the terms with equal number of $\rho_{R_p}^s$ in between two interactions, give equal contributions in magnitude. In the end we are left with only four distinct terms. These four terms multiplied by m give us the expression for R\'enyi entropy flows in Eq.(\ref{eq:renyi}).
    \section{One qubit and Two-qubit junctions}\label{app:gates}
    
    Let $\text{I}$, $\sigma^x$, $\sigma^y$ and $\sigma^z$ be the four pauli operators. The system Hamiltonian for both junctions can be written as $H_S=H_{S0}+H_{dr}$ where $H_{dr}=\Omega\cos\left(\omega_{dr}t\right)\sigma^x$ and $S=\{X,CR\}$. In the rotating frame of the drive frequency and applying RWA we get $\tilde{H}_{dr}=\frac{\Omega}{2}X$. The transition between two eigenstates of $H_{S0}$, induced by drive Hamiltonian $H_{dr}$, is the strongest when at resonance with eigenenergy difference. The transition rate 
    $$\alpha_{ij} =  |\langle v_i |H_{dr}| v_j \rangle|$$  
where $\{v_l\}$ are eigenvectors of $H_{S0}$, depends on the matrix-elements of $H_{dr}$ in diagonal frame of $H_{S0}$.    
   \subsection*{One qubit }
   X-gate is a single qubit junction used to flip the state of a qubit when applied for finite time. The bare Hamiltonian for this setup is $\text{H}_X=-\frac{\omega}{2}\hat{\sigma}^z+H_{dr}$. In the rotating frame, we have $\tilde{H}_X=\tilde{H}_{dr}$ (at resonance). The transition rate in this case is $$\alpha_{01} =\frac{\Omega}{2} \;\text{when} \;,\; \omega_{dr}=\omega.$$
\subsection*{Two qubit }
Cross-Resonance (CR) gate uses a two-qubit setup, where control qubit $q_1$ is driven at the dressed frequency of the target qubit $q_2$. It is used to create ZX rotation as a part of CNOT operation. Consider the bare Hamiltonian of this two qubit junction, $\text{H}^{\rm 2q}=-\frac{\omega_1}{2}\hat{\sigma}^z_1-\frac{\omega_2}{2}\hat{\sigma}^z_2+J\left(
\sigma_1^\dagger\sigma_2+\mathrm{H.c.}
\right)+H_{dr}$, where $H_{dr}$ is applied only on qubit $q_1$. The eigenvalues of $H_{CR0}$ are 
\begin{align*}
E_1 &= \frac{\omega_1 + \omega_2}{2} \\
E_2 &= +\sqrt{J^2 + (\Delta/2)^2} \\
E_3 &= -\sqrt{J^2 + (\Delta/2)^2} \\
E_4 &= -\frac{\omega_1 + \omega_2}{2}
\end{align*}
There are two possible resonances in this case 
\begin{align*}
\Tilde{\omega}_1 &=\frac{\omega_1 + \omega_2}{2}+\sqrt{J^2 + (\Delta/2)^2}\\
\Tilde{\omega}_2 &=\frac{\omega_1 + \omega_2}{2}-\sqrt{J^2 + (\Delta/2)^2}
\end{align*}
This is approximately where the first and last peak in Fig.\ref{fig:f_peaks} is observed. In the eigenbasis of $H_{CR0}$, the drive Hamiltonian in rotating frame becomes
\begin{equation*}\label{H_dr2}
\tilde{H}_{dr} =
\frac{\Omega}{2}
\begin{pmatrix}
0 & \cos\theta & -\sin\theta & 0 \\
\cos\theta & 0 & 0 & \sin\theta \\
-\sin\theta & 0 & 0 & \cos\theta \\
0 & \sin\theta & \cos\theta & 0
\end{pmatrix}
\end{equation*}
where $\theta = \frac{1}{2} \arctan\left(\frac{2J}{\Delta}\right)$. The transition rates in this case are 
    \begin{align}
         \alpha_{00,10}&=\alpha_{01,11}=\frac{\Omega}{2}\sin\theta \;,\; \text{when}\; \omega_{dr}=\Tilde{\omega}_1 \\
        \alpha_{00,01}&=\alpha_{10,11} =\frac{\Omega}{2}\cos\theta \;,\; \text{when}\; \omega_{dr}=\Tilde{\omega}_2 
    \end{align}
       
In the typical regime $J < \Delta$, we have that
$\sin\theta \ll \cos\theta$. Hence the peak at $\tilde{\omega}_2$ is higher than the one at $\tilde{\omega}_1$ in Fig.\ref{fig:f_peaks}. Also as $\alpha_{01}\ge \alpha_{00,10}$, drive provides more energy to in the one-qubit setup as compared to the two qubit  setup.
\section{Additional peak in the entropy flow} \label{app:anticrossings}
\begin{figure*}[ht]
    \centering
    \includegraphics[width=0.9\linewidth]{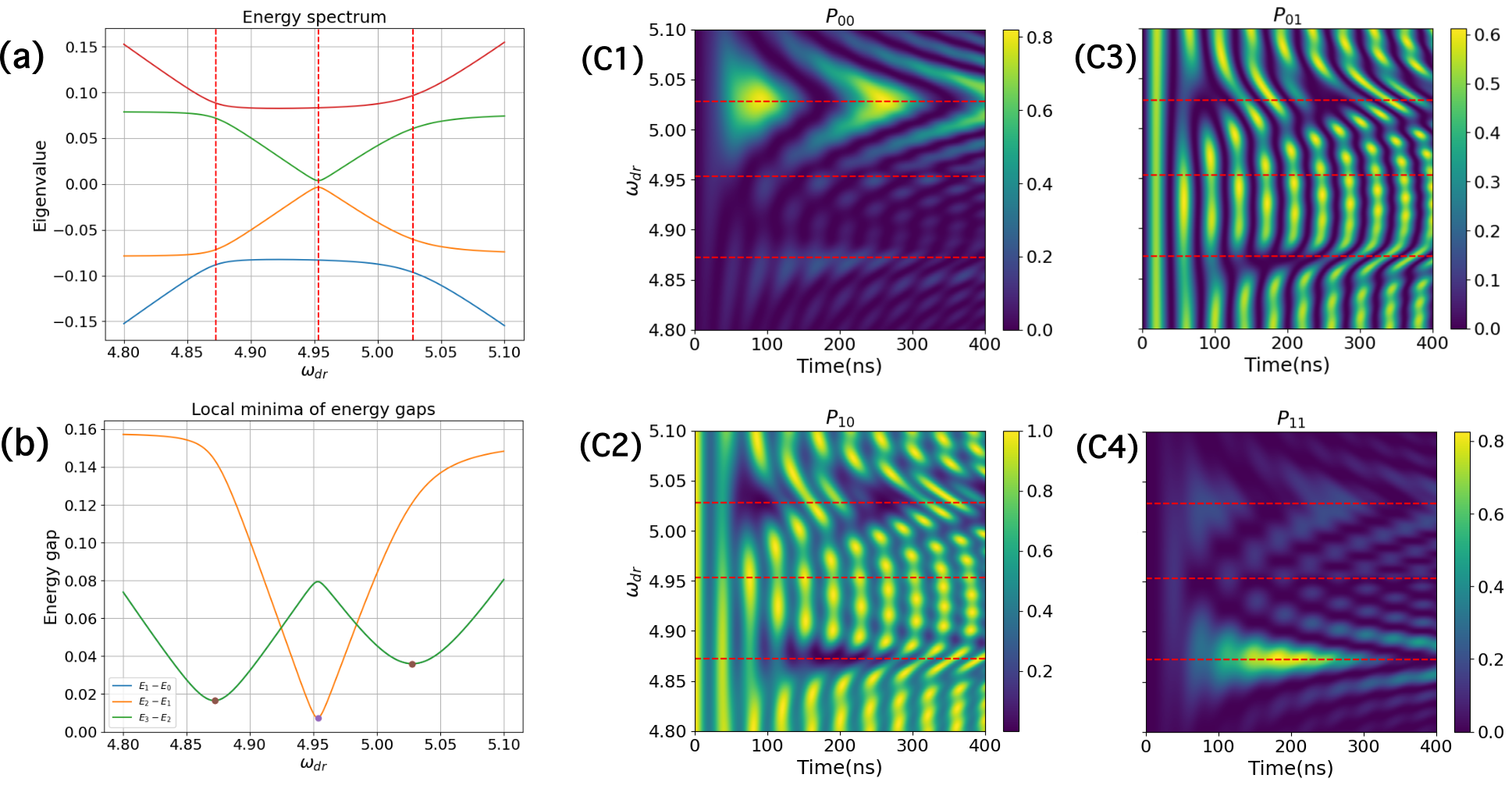}
    \caption{ \textbf{(a)Energy spectrum :} Four eigenvalues are obtained as a function of drive frequency $\omega_{dr}$ \textbf{(b)Anticrossings:} Evaluating the local minimas in the eigenenergy diagram gives the three anticrossing points. \textbf{(C1-C4) Population evolution :}  Four populations are evolved for finite time. Irregularities are seen at the same three drive frequencies where anticrossing is observed.}
    \label{fig:f_eig_evol0}
\end{figure*}

The two peaks in the entropy flows in Fig.\ref{fig:f_peaks} are already explained in Appendix \ref{app:gates} by treating $H_{dr}$ as a perturbation. Instead, if we now find the spectrum of the whole Hamiltonian $\tilde{H}_S$ in rotating frame of the drive frequency, the middle peak can also be explained. For two qubit  setup we have
\begin{align}
\tilde{H}^{\rm 2q}
&=
-\frac{\omega_1}{2}\hat{\sigma}^z_1
-
\frac{\omega_2}{2}\hat{\sigma}^z_2
+
\frac{\Omega}{2}\cos(\omega_{\rm dr}t)\,\hat{\sigma}^x_1
\nonumber\\
&+J\left(
\sigma_1^\dagger\sigma_2+\mathrm{H.c.}
\right),
\label{eq:2qH}
\end{align}
    Eigenvalues of this Hamiltonian as a function of drive frequency $\omega_{dr}$ are given in Fig.\ref{fig:f_eig_evol0}\blu{(a)}. The values of $\omega_{dr}$ at which three anticrossings appear as shown in Fig.\ref{fig:f_eig_evol0}\blu{(b)} is where the eigenenergies come closest and energy exchange becomes easier. Fig.\ref{fig:f_eig_evol0}\blu{(C1-C4)}  show the population evolutions for varying $\omega_{dr}$. A distinct behaviour is obeserved at the three anticrossings. These are the values of $\omega_{dr}$ at which we observe three peaks in Fig.\ref{fig:f_peaks}.
\end{document}